\documentclass[epj,referee]{svjour}
\usepackage{amsfonts}
\usepackage{amssymb}
\usepackage{latexsym}
\usepackage{graphics}
\usepackage{dsfont}
\usepackage{array}
\usepackage{tabularx}

\begin{document}
\title{Analysis on the evolution process of
BFW-like model with explosive percolation of multiple giant
components}
\author{Renquan Zhang, Wei Wei\thanks{\emph{e-mail:} weiw@buaa.edu.cn}, Binghui Guo,
Yang Zhang \and Zhiming Zheng }

\institute{LMIB and School of Mathematics and Systems Science,
Beihang University, Beijing, P.R.China, 100191}
\date{Received: date / Revised version: date}
%
\abstract{Recently, the modified BFW model on random graph [Phys.
Rev. Lett., 106, 115701 (2011)], which shows a strongly
discontinuous percolation transition with multiple giant components,
has attracted much attention from physicists, statisticians and
materials scientists. In this paper, by establishing the theoretical
expression of evolution equations on the modified BFW model, the
steady-state and evolution process are analyzed and a close
correspondence is built between the values of parameter $\alpha$ and
the number of giant components in steady-states, which fits very
well with the numerical simulations. In fact, with the value of
$\alpha$ decreasing to $0.25$, the error between theoretical and
numerical results is smaller than $4\%$ and trends to $0$ rapidly.
Furthermore, the sizes of giant components for different evolution
strategies can also be obtained by solving some constraints derived
from the evolution equations. The analysis of the steady-state and
evolution process is of great help to explain why the percolation of
modified BFW model is explosive and how explosive it is. }
\maketitle
\section{Introduction}
\label{intro} Percolation is a classical model in statistical
physics, probability theory, materials science, complex networks and
epidemiology, which is initiated as a mathematical framework for the
study of random physical processes such as flow through a disordered
porous medium. The research in percolation is not only of academic
interest but also of considerable practical value. During the last
five decades, percolation theory has found a broad range of
application in epidemic spreading, porous media, robustness of
networks to attacks, etc
\cite{percolationtheory,percolationapplication,amaral}.

Percolation has been studied on various topological structures such
as scale-free network, lattices with different dimensional, random
graph, etc. Taking percolation model on the
Erd$\ddot{o}$s-R$\acute{e}$nyi random graph (ER model) as an
example, this model is one of the most simple and classical models
that undergo a phase transition of a emerging giant component.
Typically, percolation phase transition is considered as a robust
second-order transition until a recent work by D. Achlioptas, R. M.
D'Souza, and J. Spencer \cite{Achlioptas}, in which they propose
that the phase transition of some certain Achlioptas process is
discontinuous and call it {\it explosive percolation}. This interest
phenomenon leads to intensive studies on the other models like
scale-free network \cite{scale-free1,scale-free2}, local cluster
aggregation model \cite{aggregation} and lattices
\cite{lattices1,lattices2}. More recently, it has been demonstrated
that all Achlioptas processes have continuous phase transitions in
the mean-field limit
\cite{discontinuous1,discontinuous2,discontinuous3,discontinuous4,discontinuous5,discontinuous6}
. But some other kinds of models, which have different and special
rule of evolution, such as triangle rule \cite{d1}, largest cluster
rule \cite{d2}, etc \cite{d3,d4,d5,d6,d7,d8,d9,d10}, have been
analyzed in details and indeed exhibit explosive percolation.

In particular, the BFW model on random graphs, originally introduced
by Bohman, Frieze, and Wormald \cite{BFW}, is similar to Achlioptas
processes but more restricted. The recent work of W. Chen and R. M.
D'Souza \cite{chen1,chen2}, shows a strongly explosive percolation
with multiple giant components in BFW model. It is also shown that
with smaller parameter values, the transition will become more {\it
explosive} and the number of giant components will increase.
Furthermore, K. J. Schrenk {\it et al.} \cite{latticenew} generalize
the results to the lattice with different dimensions.

So far, although one-dimensional and mean-field percolations have
been solved theoretically, the others still remain on researching,
especially the explosive percolation, which attracts much attention
from physicists, statisticians and materials scientists. As a
typical member of explosive percolation, the percolation threshold
of BFW model has been analyzed, but many other properties are still
not clear, which drives us to investigate the evolution process of
BFW model with both simulation and theoretical method.

This paper is organized as follows. In section 2, we introduce the
BFW algorithm with parameter $\alpha$ in details; the mathematical
expression of BFW model is established and analyzed in theory, by
which, we obtain the steady-state condition and evolution regulation
of BFW model for any $\alpha$. In section 3, by analyzing the
evolution procedure of BFW model, we find the relationship between
parameter $\alpha$ and the steady-state, that is when
$\alpha\in(\frac{1}{m+1},\frac{1}{m}]$, BFW algorithm must stabilize
with $m$ giant components, for any $m\in\mathbb{N}^+$. Furthermore,
size of these components must satisfy some constraint equations
which are given in our paper.

\section{Dynamical behaviors of BFW($\alpha$) model}

The BFW model on random graph is firstly introduced by T. Bohman, A.
Frieze, and N. C. Wormald \cite{BFW}, aiming to choose a subset
$A\in\{e_1, e_2, ..., e_{2t}\}$ with $|A|=t$ such that for $t$ as
large as possible the size of the largest component in $G=(n,A)$ is
$o(n)$ (i.e. $G$ does not contain a giant component); here $n$
denotes the number of nodes and $\{e_1,e_2, ...\}$ are the sequence
of edges chosen uniformly at random from the edge set of complete
graph; $A$ represents the set of accepted edges (initialized to
$A=\emptyset$) and $t=|A|$ represents the number of accepted edges.

According to the BFW model, one of the sampled edges is considered
at each step, and either accepted to the graph or rejected provided
that the fraction of accepted edges is never smaller than the
decreasing function $g(k)$, which is asymptotically approaching the
value $1/2$. If taking $u$ as the total number of sampled edges, the
fraction of accepted edges is represented by $t/u$; $k$ denotes the
stage and the function $g(k)=1/2+1/\sqrt{2k}$. This model is much
similar to Achlioptas process and shows that the threshold of a
giant component is $t=c^*n$ where $c^*$ satisfies a certain
transcendental equation and $c^*\in [0.9792,0.9793]$. This result
has been verified by theoretical methods \cite{BFW} and simulations
\cite{chen1}.

Recently, the BFW model is extended to BFW($\alpha$) one and
analyzed by W. Chen and R. M. D'Souza \cite{chen1} with modifying
the function $g(k)$ to $\alpha+1/\sqrt{2k}$. It is shown that
multiple giant components appear in a strongly explosive percolation
transition. Furthermore, with the value of $\alpha$ decreasing, the
threshold will delay and the phase transition will be more
explosive. In the following sections, we will discuss the reason and
properties of these phenomenons and provide theoretical analysis.

For the theoretical analysis on the BFW($\alpha$) model, the
\emph{BFW($\alpha$) algorithm} is shown as follows:

\quad \\
\noindent
\begin{tabular}[c]{p{480pt}l}
\textbf{algorithm}  \textbf{BFW($\alpha$)} \\
1\ \ \ \textbf{begin}\\
2\ \ \ \ \ \ \ \ \ \ \ \texttt{$A=\emptyset$;}\\
3\ \ \ \ \ \ \ \ \ \ \ \texttt{$k=2$;}\\
4\ \ \ \ \ \ \ \ \ \ \ \texttt{$t=u=1$;}\\
5\ \ \ \ \ \ \ \ \ \ \ \texttt{$e$ is a randomly sampled edge;}\\
6\ \ \ \ \ \ \ \ \ \ \ \textbf{while}(\texttt{$t<2n$})\\
7\ \ \ \ \ \ \ \ \ \ \ \textbf{begin}\\
8\ \ \ \ \ \ \ \ \ \ \ \ \ \ \ \ \ \ \ \texttt{$l$= Maximum size of component in $A\bigcup\{e\}$;}\\
9\ \ \ \ \ \ \ \ \ \ \ \ \ \ \ \ \ \ \ \textbf{if}(\texttt{$l\leq k$})\\
10\ \ \ \ \ \ \ \ \ \ \ \ \ \ \ \ \ \ \ \ \ \ \ \ \ \ \ \texttt{$A=A\bigcup\{e\}$;}\\
11\ \ \ \ \ \ \ \ \ \ \ \ \ \ \ \ \ \ \ \ \ \ \ \ \ \ \ \texttt{$t=t+1$;}\\
12\ \ \ \ \ \ \ \ \ \ \ \ \ \ \ \ \ \ \ \ \ \ \ \ \ \ \ \texttt{$u=u+1$;}\\
13\ \ \ \ \ \ \ \ \ \ \ \ \ \ \ \ \ \ \ \ \ \ \ \ \ \ \ \texttt{sample an edge $e$ randomly;}\\
14\ \ \ \ \ \ \ \ \ \ \ \ \ \ \ \ \ \ \ \textbf{else if}(\texttt{$t/u<\alpha+1/\sqrt{2k}$})\\
15\ \ \ \ \ \ \ \ \ \ \ \ \ \ \ \ \ \ \ \ \ \ \ \ \ \ \ \texttt{$k=k+1$;}\\
16\ \ \ \ \ \ \ \ \ \ \ \ \ \ \ \ \ \ \ \textbf{else}\\
17\ \ \ \ \ \ \ \ \ \ \ \ \ \ \ \ \ \ \ \ \ \ \ \ \ \ \ \texttt{$u=u+1$;}\\
18\ \ \ \ \ \ \ \ \ \ \ \ \ \ \ \ \ \ \ \ \ \ \ \ \ \ \ \texttt{sample an edge $e$ randomly;}\\
19\ \ \ \ \ \ \ \ \ \ \ \textbf{end}\\
20\ \ \ \textbf{end}\\
\end{tabular}

\subsection{Evolution analysis on the BFW($\alpha$) model}

To analyze the evolution of BFW($\alpha$) model and its
steady-states, we consider the following variables: $k$, $t$, $u$
and $n$ possess the same meaning as they are in the BFW($\alpha$)
algorithm; $m$ represents the number of components; $C_i$ denotes
the fraction of $i$th largest component.

In BFW($\alpha$) algorithm, there are three cases when an edge is
sampled:

$\bullet${\it Case \uppercase\expandafter{\romannumeral1}}: the
vertices of sampled edge are in the same component;

$\bullet${\it Case \uppercase\expandafter{\romannumeral2}}: they are
in two components $C_i$ and $C_j$ and $C_i+C_j<k/n$;

$\bullet${\it Case \uppercase\expandafter{\romannumeral3}}: they are
in two components $C_i$ and $C_j$ and $C_i+C_j>k/n$.

According to the BFW($\alpha$) algorithm, we sample a random edge at
step $u$: in Case \uppercase\expandafter{\romannumeral1}, the edge
is also accepted; in Case \uppercase\expandafter{\romannumeral2},
the edge is also accepted and two components $C_i$ and $C_j$ merge
together; in Case \uppercase\expandafter{\romannumeral3}, we should
consider the constraint condition $t/u<\alpha+1/\sqrt{2k}$ (on the
14th line of the BFW($\alpha$) algorithm). This constraint condition
is the kernel hard core of BFW($\alpha$) model, which ensures either
components are increasing evenly or dramatically.

Let's first introduce a function $f_\alpha(t,u,k)$, which denotes
the maximum acceptable value of $\triangle k$ at one step. Due to
the BFW($\alpha$) algorithm, if the rate of accepted edges $t/u$ is
smaller than $\alpha$, any sampled edge should be accepted; else,
$k$ can only increase until the condition $t/u<\alpha+1/\sqrt{2k}$
is invalid. Thus, the function $f_\alpha(t,u,k)$ is shown as
follows:

\begin{equation}
f_\alpha(t,u,k)= \left\{
\begin{array}{ll}
\min\{x\in \mathbb{N}^+|\frac{t}{u}\ge\alpha+\frac{1}{\sqrt{2(k+x)}}\}, &\mbox{if}\  \frac{t}{u}>\alpha \\
\infty,  &\mbox{if}\  \frac{t}{u}\leq\alpha.
\end{array}
\right.
\end{equation}

According to the definition of $f_\alpha$, when a randomly edge is
sampled between two components $C_i$ and $C_j$, if and only if
$f_\alpha\geq n(C_i+C_j)-k$, we can accept the edge ($t\leftarrow
t+1$) and the components $C_i$ and $C_j$ merge together
($m\leftarrow m-1$). Moreover, $k$ can change by no more than
$n(C_i+C_j)-k$ and $f_\alpha(t,u,k)$, so we have that in one step:

\begin{eqnarray}
&&\triangle k=\min\left(n(C_i+C_j)-k,f_\alpha\right),\\
&&\triangle t=-\triangle m=\delta\left(n(C_i+C_j)-k,f_\alpha\right).
\end{eqnarray}
Here $\delta(x,y)=1$ if $x\leq y$ and $\delta(x,y)=0$ otherwise.

As $u$ increases, the evolution equations of $k$, $t$ and $m$ are
established as follows:

\begin{eqnarray}
&&\frac{\mathrm{d}k}{\mathrm{d}u}=2\sum_{C_i+C_j>k/n}\min\left(n(C_i+C_j)-k,f_\alpha\right)C_iC_j,\\
&&\frac{\mathrm{d}m}{\mathrm{d}u}=-P_{2}(t,u,k)\nonumber\\
&&\ \ \ \ \ \ -2\sum_{C_i+C_j>k/n}\delta\left(n(C_i+C_j)-k,f_\alpha\right)C_iC_j,\\
&&\frac{\mathrm{d}t}{\mathrm{d}u}=P_{1}(t,u,k)+P_{2}(t,u,k)\nonumber\\
&&\ \ \ \ \ \ +2\sum_{C_i+C_j>k/n}\delta(n(C_i+C_j)-k,f_\alpha)C_iC_j.
\end{eqnarray}
Here the function $P_{1}(t,u,k)$ is defined as the probability that
the vertices of a randomly sampled edge at step $u$ are in the same
component (Case \uppercase\expandafter{\romannumeral1}); similarly,
the function $P_{2}(t,u,k)$ is defined as the probability that they
are in two components with sum smaller than $k$ (Case
\uppercase\expandafter{\romannumeral2}). Therefore, we can simply
obtain:

\begin{eqnarray}
&&P_{1}(t,u,k)=\sum_{i=1}^m C_i^2,\\
&&P_{2}(t,u,k)=2\sum_{C_i+C_j\leq k/n}C_iC_j.
\end{eqnarray}

For Eq.(4), $k$ is the upper bound of size of the largest component
and never changes in Case \uppercase\expandafter{\romannumeral1} and
\uppercase\expandafter{\romannumeral2}; only in Case
\uppercase\expandafter{\romannumeral3}, $k$ can change by no more
than $n(C_i+C_j)-k$ and $f_\alpha(t,u,k)$, i.e., $\triangle
k=\min\left(n(C_i+C_j)-k,f_\alpha(t,u,k)\right)$. For Eq.(5),
$\triangle m=0$ in Case \uppercase\expandafter{\romannumeral1} and
$-1$ in Case \uppercase\expandafter{\romannumeral2} respectively; in
Case \uppercase\expandafter{\romannumeral3}, the number of
components will decrease by 1 if and only if $f_\alpha\geq
n(C_i+C_j)-k$, i.e., $\triangle
m=-\delta\left(n(C_i+C_j)-k,f_\alpha\right)$. For Eq.(6), the
sampled edge must be accepted in Case
\uppercase\expandafter{\romannumeral1} and
\uppercase\expandafter{\romannumeral2}, so $\triangle t=1$; similar
to the $m$ of Eq.(5) in Case \uppercase\expandafter{\romannumeral3},
edge can be accepted when $f_\alpha\geq n(C_i+C_j)-k$ and $\triangle
t=\delta\left(n(C_i+C_j)-k,f_\alpha\right)$.

\subsection{Steady-state conditions of evolution of the BFW($\alpha$) model}

Although the Eq.(4)-(6) are unsolvable, some interest properties and
results can still be deduced from these equations, especially the
steady-state conditions.

Taking the right side of Eq.(4) and (5) as $0$, we can obtain
$P_{2}(t,u,k)=0$, $\min\left(n(C_i+C_j)-k,f_\alpha\right)=0$ and
$\delta\left(n(C_i+C_j)-k,f_\alpha\right)=0$. Furthermore, for two
components with sum smaller than $k$, they must merge together, but
merging operation is forbidden after system stabilizes, so any two
giant components stay with $n(C_i+C_j)-k>0$ after stabilizing, then
the steady-state conditions can be simplified to be:

\begin{equation}
\left\{
\begin{array}{l}
f_\alpha(t,u,k)=0\\
P_{2}(t,u,k)=0.
\end{array}
\right.
\end{equation}

According to the definition, $f_\alpha(t,u,k)=0$ if and only if
$t/u>\alpha+1/\sqrt{2k}$, so we just need to prove
$t/u>\alpha+1/\sqrt{2k}$ and $P_{1}(t,u,k)>\alpha$ are equivalent.

For $P_{1}(t,u',k)>\alpha$ with some $u'$, we can prove that
$t/u>\alpha+1/\sqrt{2k}$ for any $u>u'$. Doing calculations on both
sides of Eq.(6) from an initial state $(t_{0},u_{0})$ to a current
state $(t_{1},u_{1})$, we have:

\begin{eqnarray}
t_{1}-t_{0} && =\int_{u_{0}}^{u_{1}} P_{1}\, du\nonumber\\
&&\ \ +\int_{u_{0}}^{u_{1}} 2\sum_{C_i+C_j>k/n}\delta(n(C_i+C_j)-k,f_\alpha)C_iC_j\, du\nonumber\\
&& >\int_{u_{0}}^{u_{1}} P_{1}\, du.\nonumber
\end{eqnarray}

No matter how large $u$ is, we can always find some $u_{0}>u$ with
$t_{0}/u_{0}>\alpha$ because BFW($\alpha$) model ensures the
fraction of accepted edges never smaller than $\alpha$. Supposed
$t/u$ is always smaller than $\alpha$, by the rule of BFW($\alpha$)
model, the sampled edge must be accepted and $t$, $u$ increase
accordingly in each step, which will lead to the increase of the
value of $t/u$ and finally make $t/u>\alpha$.

For $P_{1}$, only when a sampled edge linking $C_i$ and $C_j$ is
accepted, the part of $C_i^2+C_j^2$ changes to $(C_i+C_j)^2$, which
will make the value of $P_{1}(t,u,k)$ increase; otherwise,
$P_{1}(t,u,k)$ will never change. So once $P_{1}(t,u,k)>\alpha$ for
some $u$, it will be kept for ever.

In summary, once $P_{1}(t,u',k)>\alpha$ for some $u'$, choosing
$u_{0}>u'$ with $t_{0}/u_{0}>\alpha$, we obtain:

\begin{eqnarray}
t_{1}>t_{0}+\alpha\int_{u_{0}}^{u_{1}}\, du>\alpha u_{1}. \nonumber
\end{eqnarray}
Notice that the formula above is correct for any $u_{1}>u_{0}$.
After the giant components come up, we have:

\begin{eqnarray}
k\sim nC_{max} \sim O(n). \nonumber
\end{eqnarray}
So $1/\sqrt{2k} \sim o(1)$ and it can be ignored when $n$ is large
enough. Therefore, when $P_{1}(t,u_{1},k)>\alpha$ for any
$u_{1}>u'$, we obtain $t_{1}/u_{1}>\alpha+1/\sqrt{2k}$.

In the other side, if $f_\alpha(t,u,k)=0$ keeps for any $u>u'$,
Eq.(6) turns to be:

\begin{eqnarray}
\frac{\mathrm{d}t}{\mathrm{d}u}=P_{1}(t,u,k). \nonumber
\end{eqnarray}
In order to ensure $t/u>\alpha+o(1)(\forall u>u')$, we need the
slope $P_{1}(t,u,k)>\alpha$.$\square$

Finally, we can obtain the steady-state conditions for any $u>u'$:

\begin{equation}
\left\{
\begin{array}{l}
P_{1}(t,u,k)>\alpha\\
P_{2}(t,u,k)=0.
\end{array}
\right.
\end{equation}
In fig.1, simulations have verified this conclusion. For different
values of $\alpha$, the values of $t/u$ and $k/n$ change until
$P_{1}(t,u,k)=\sum_{i=1}^{100} C_i^2>\alpha$, which means the
BFW($\alpha$) system evolves until Eq.(10) taking effect.

\begin{figure}
\centering \resizebox{1\columnwidth}{!}{
\includegraphics{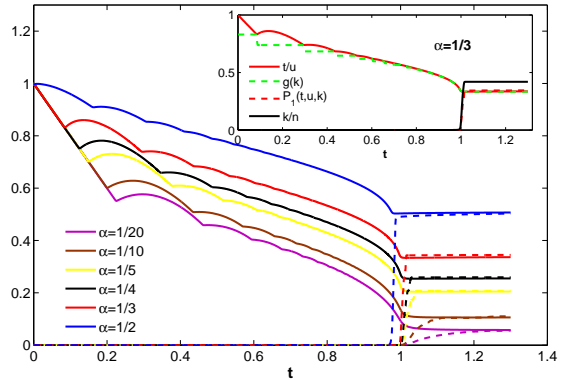}
}
\caption{Simulation of BFW($\alpha$) model by 100 random instances
with $N=10^6$ nodes. The solid lines denote the fraction of accepted
edges for different values of $\alpha$, which can be calculated by
$t/u$ in algorithm; the dashed lines denote
$P_{1}(t,u,k)=\sum_{i=1}^{100} C_i^2$. The above subgraph shows the
dynamical evolution of variables $k/n$, $t/u$ and $P_{1}(t,u,k)$
with $\alpha=1/3$. By the results, when $P_1(t,u,k)>\alpha$, $t/u$
keeps greater than $\alpha$ and $k$ keeps still. }
\label{fig:1}       
\end{figure}

\subsection{Merging mechanism on the multiple giant components of BFW($\alpha$) model}

Since in Case \uppercase\expandafter{\romannumeral1} and
\uppercase\expandafter{\romannumeral2}, sampled edges are always
accepted; we just need to explicitly consider the Case
\uppercase\expandafter{\romannumeral3}. Considering
$t/u\geqslant\alpha+1/\sqrt{2k}$ at step $u$; according to the
BFW($\alpha$) algorithm, when the edge $e_{u+1}$ is sampled at step
$u+1$ and $t/(u+1)\geqslant\alpha+1/\sqrt{2k}$, the edge $e_{u+1}$
is simply rejected. But if $t/(u+1)<\alpha+1/\sqrt{2k}$, $k$ needs
to increase. To obtain the maximum accepted change of $k$, i.e.
$f_\alpha$, we set $t/u$ to be the smallest value:

\begin{eqnarray}
\frac{t}{u}=\alpha+\frac{1}{\sqrt{2k}},\nonumber
\end{eqnarray}
differentiating $u$ on both sides by $k$ we find that:

\begin{eqnarray}
\frac{\mathrm{d}u}{\mathrm{d}k}=\frac{1}{2\sqrt{2}\left(\alpha+\sqrt{1/2k}\right)^2}\
\frac{t}{k^{3/2}}.
\end{eqnarray}

Before the giant component appears, we consider that $t\sim O(n)$,
$k\sim nC_i\sim o(n)$. With $\mathrm{d}u=1$ and Eq.(11), we obtain
$f_\alpha\sim \mathrm{d}k\sim O(k^{\frac{3}{2}}/n)$. Let $\mathcal
S$ denote the component set $\left\{C_i\ |\ k/2<C_in<k\right\}$. If
the sampled edge links components $C_i$ and $C_j$, where
$C_i,C_j\in\mathcal S$, then $n(C_i+C_j)-k\sim O(k)\gg
O(k^{\frac{3}{2}}/n)$. That leads to
$\delta\left(n(C_i+C_j)-k,f_\alpha\right)=0$, which means this edge
is rejected. So the edge linking two components of $\mathcal S$ must
be rejected and only the edge linking to at least one component of
$\complement\mathcal S$ can be received. For the process is
performed successively, either a new member in $\mathcal S$ comes up
or the scale of an original one in $\mathcal S$ becomes more close
to $k$. That is the key for coexisting multiple giant components,
and they are expected to grow simultaneously before a critical
point(fig.2).

\begin{figure}
\centering \resizebox{1\columnwidth}{!}{
\includegraphics{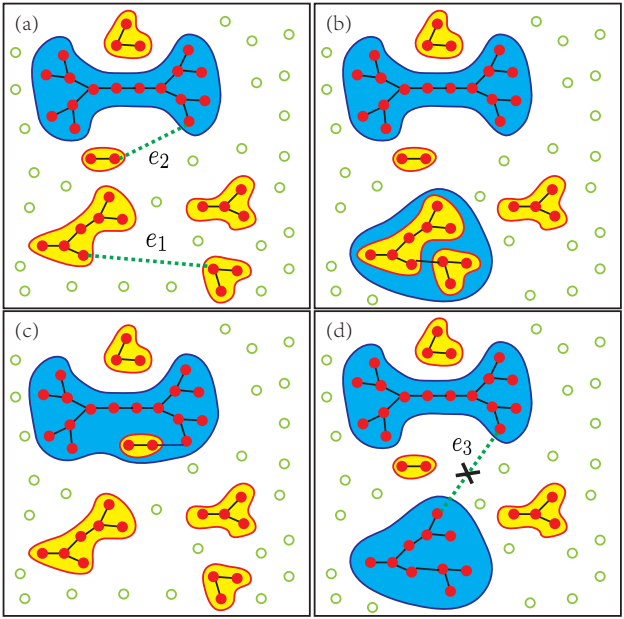}
}
\caption{An example for the evolution of BFW($\alpha$) model. The
blue (yellow) components are (not) in the set $\mathcal S$. The
green cycles denote the isolated nodes. Subgraph $(a)$ provides the
initial graph, in which two random edges $(e_1, e_2)$ are picked in
each step yet only one is added to the graph based on selection rule
of BFW($\alpha$) model, whereas the other is discarded. Subgraph
$(b)$ is the situation of adding edge $e_1$, two tiny components
merge into a giant one. Subgraph $(c)$ is the situation of adding
edge $e_2$, a tiny component merges into a giant one, forming a new
member of set $\mathcal S$. In subgraph $(d)$, the sampled edge
$e_3$ linking two giant components is rejected.} \label{fig:2}
\end{figure}

Similarly, after the giant components appear, the $k\sim nC_i\sim
O(n)$,$\forall C_i\in\mathcal S$, so we have:

\begin{eqnarray}
f_\alpha(t,u,k)\sim O(n^{1/2})\ll n(C_i+C_j)-k.
\end{eqnarray}
Then we still have $\delta\left(n(C_i+C_j)-k,f_\alpha\right)=0$ and
the members of $\complement\mathcal S$ keep merging into $\mathcal
S$ until $P_{2}=0$. According to the steady-state conditions
Eq.(10), if $C_i$ satisfies $P_{1}>\alpha$, the system will
stabilize; else, $P_1(t,u,k)$, which is the probability that the
vertices of sampled edge are in the same component, is smaller than
$\alpha$. As it is proved above, if $P_{1}<\alpha$, we can always
have some $u$ with $t/u<\alpha+1/\sqrt{2k}$, which makes $k$ keep
increasing (the $14th$ and $15th$ lines of the algorithm) until two
components merge together.

Furthermore, only two minimum components (marked as
$C_1^{min},C_2^{min}$) can merge together. We define $P$ as the
probability of any other two components (marked as $C_i,C_j$)
merging together before the two minimum ones, then

\begin{eqnarray}
P\leqslant\left(1-2C_1^{min}C_2^{min}\right)^{n(C_i+C_j-C_1^{min}-C_2^{min})\big/\overline{f_\alpha(t,u,k)}}.
\end{eqnarray}
For one step, $\triangle
k=\min\left(n(C_i+C_j)-k,f_\alpha\right)=f_\alpha$. Let's take
$\overline{f_\alpha(t,u,k)}$ as average increase of $k$ for one
step. Based on Eq.(12), $\overline{f_\alpha(t,u,k)}\sim O(n^{1/2})$.
When $k$ increases to be larger than $n(C_1^{min}+C_2^{min})$ but
smaller than $n(C_i+C_j)$, that only one edge linking $C_1^{min}$
and $C_2^{min}$ is sampled can make them merge together. So in order
to ensure that $C_1^{min}$ and $C_2^{min}$ can't merge together
before $C_i$ and $C_j$, we need $k$ to increase by
$n(C_i+C_j-C_1^{min}-C_2^{min})$ without any components merging,
which means
$n(C_i+C_j-C_1^{min}-C_2^{min})\big/\overline{f_\alpha(t,u,k)}$
edges not linking $C_1^{min}$ and $C_2^{min}$ should be added. So
the probability $P$ satisfies Eq.(13).

As the system size $n\to\infty$, the number of needed edges
$n(C_i+C_j-C_1^{min}-C_2^{min})\big/\overline{f_\alpha(t,u,k)}\to
O(n^{1/2})\to\infty$, causes $P\to 0$, which means two minimum
components can merge before any other two components. This
phenomenon can also be verified by simulation of BFW($\alpha$) model
(fig.3).

\begin{figure}
\centering \resizebox{1\columnwidth}{!}{
\includegraphics{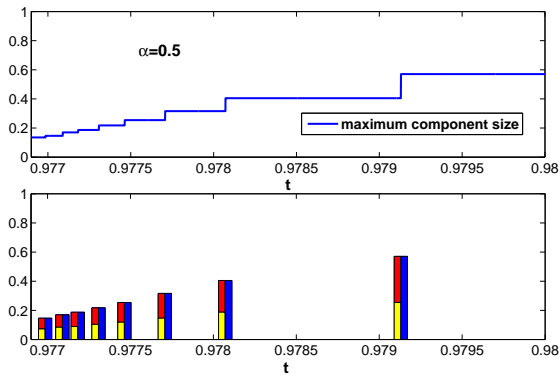}
}
\caption{ Up: The maximum component's size in critical interval of
the BFW($\alpha$) process; the horizontal axis $t$ reflects a "jump
point" at which the maximum component increases dramatically. Down:
the yellow and red rectangles represent
 the first and second minimum components before "jump point",
 and the blue ones represent the maximum component after "jump point". At the "jump point",
 the two minimum components emerge into the maximum one. All the other components in $\mathcal S$
 keep unchanged at "jump point". }
\label{fig:3}
\end{figure}

\section{Quantitative properties on the giant components of BFW($\alpha$) model}

Since we have analyzed the BFW($\alpha$) model with fixed $\alpha$
in detail, in this section, we are going to calculate the number and
size of giant components with arbitrary $\alpha$ in theory. Taking
$C_i^m(\alpha)$ as the $i$th largest giant component of steady-state
with $m$ giant components ({\it$m$-steady-state}) to replace $C_i$
above, we will find some common properties of $C_i^m(\alpha)$ when
$\alpha$ belongs to some intervals.

\subsection{Steady states with different evolution parameter $\alpha$}

In BFW($\alpha$) model, parameter $\alpha$ plays a key role on the
problems when the system can stabilize and which state the system
can stabilize in. Defining $(\alpha_{m+1},\alpha_m]$ as the
$m$-steady-state interval in which there exists $m$ giant
components, $m=1,2,3 \cdot\cdot\cdot$, for any $\alpha$ with
$\alpha_{m+1}>\alpha\geqslant\alpha_m$, the value of $C_i^m(\alpha)$
are all the same when system stabilizes (fig.4), so we take $C_i^m$
instead of $C_i^m(\alpha)$ briefly for all
$\alpha\in(\alpha_{m+1},\alpha_m]$. Thus, the members in set
$\mathcal S$ evolve similarly with different phases in different
intervals of $\alpha$.

\begin{figure}
\centering \resizebox{1\columnwidth}{!}{
\includegraphics{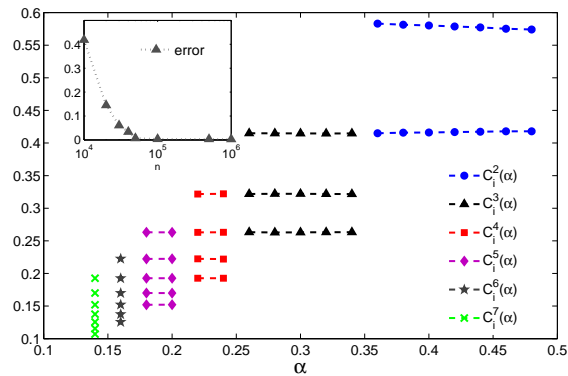}
}
\caption{Size of giant components in set $\mathcal S$ with different
$\alpha$. cycles, triangles, squares, diamonds, stars and crosses
denote the fraction of giant components in steady state of $m=2,
3..., 7$ respectively. The results are obtained by 100 random
instances with $10^6$ nodes. The above subgraph denotes the scaling
of the error of all $\mid C_i^{m+1}-C_{i+1}^m \mid$ for every
$m=2,3...,7$ and $i\in[1,m-1]$. Here the $C_i^m$ is average size of
$ith$ largest component in {\it$m$-steady-state} by simulations.}
\label{fig:4}
\end{figure}

Moreover, if system $BFW(\alpha)$ with $\alpha\in
(\alpha_{m+1},\alpha_{m}]$ has $m+1$ components $C_i^{m+1}(\alpha)$
in $\mathcal S$, there must be $P_{1}(t,u,k)<\alpha$, which leads to
its collapse and a steady phase of {\it$m$-steady-state}. As
mentioned above, Eq.(13) ensures that only the two minimum
components can merge before system stabilize and they merge to the
largest one in {\it$m$-steady-state} (fig.3). So the $C_i^m$ and
$C_i^{m+1}$ must satisfy:

\begin{eqnarray}
C_{i+1}^m=C_i^{m+1}, \forall i=1,2,...,m-1.
\end{eqnarray}

In addition, when two components merge together, $P_{1}(t,u,k)$ will
"jump" by
$(C_{m+1}^{m+1}+C_m^{m+1})^2-(C_{m+1}^{m+1})^2-(C_{m}^{m+1})^2=2C_m^{m+1}C_{m+1}^{m+1}$.
Notice that in {\it$m$-steady-state}, $P_{1}(t,u,k)$ keeps unchanged
and is larger than $\alpha\in(\alpha_{m+1},\alpha_m]$, so
$P_{1}(t,u,k)$ must be the upper bound of $\alpha$ in
{\it$m$-steady-state}:

\begin{eqnarray}
\sum_{i=1}^m (C_i^m)^2=\alpha_m.
\end{eqnarray}

Suppose all components' size in $\mathcal S$ are very close and
$\sum_{i=1}^m C_i\simeq 1$, we have theoretical expression of the
$\alpha_m$:

\begin{eqnarray}
\alpha_m=\sum_{i=1}^m (C_i^m)^2=\sum_{i=1}^m
\frac{1}{m^2}=\frac{1}{m}.
\end{eqnarray}
As the value of $\alpha$ goes smaller, the assumption is more close
to the truth by numerical results (fig.5).

\begin{figure}
\centering \resizebox{1\columnwidth}{!}{
\includegraphics{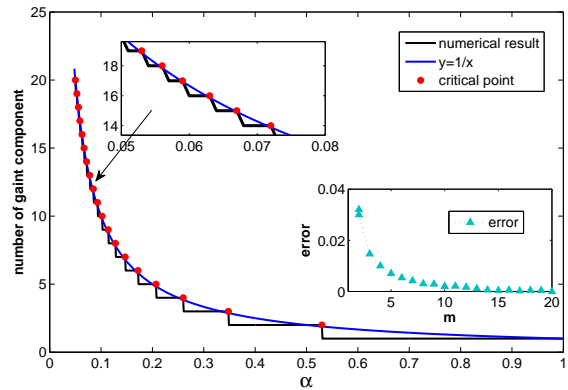}
}
\caption{Critical points of phase transition of stable giant
components' number for $\alpha\in[0.05,1]$. The red solid cycles and
black lines represent numerical results of 100 random instances with
$10^6$ nodes. The blue lines denote theoretical results. The solid
triangles in below subgraph denote the error between numerical and
theoretical results.} \label{fig:5}
\end{figure}

\subsection{Number and sizes of multiple giant components}

 Firstly, we can take the whole set $\mathcal S$ as a component and
 assume it grows similarly as the giant component on Erd$\ddot{o}$s-R$\acute{e}$nyi random
 graph, in which the number of added edges is expected to be the sampled edge number $u$. As rejected edges are almost between two components
 of $\mathcal S$, the size of whole set $\mathcal S$ is almost unchanged if we take these rejected edges
 on. According to the method of generating function \cite{newman}, the fraction $x$ of giant component satisfies the equation $1-x=e^{-2xu/n}$, here $u$ represents the
number of added edges.
 At the critical point, the threshold $t_c\simeq 1$ and $t_c/u\simeq\alpha$ when $u$ is large enough (fig.1).
 With Eq.(15), we obtain the general equations
 of the fraction of giant components $C_i^m$ for any integer $m$:

\begin{equation}
\left\{
\begin{array}{l}
\sum_{i=1}^m (C_i^m)^2=\alpha_m\\
\sum_{i=1}^m C_i^m=x_m\\
1-x_m=e^{-2x_m/\alpha}.
\end{array}
\right.
\end{equation}
Here $x_m$ denotes the fraction of whole set $\mathcal S$ which has
$m$ giant components. Notice for case $m=2$, Eq.(17) can be solved
uniquely (due to error of $\alpha_m$ in fig.5, the accurate result
$\alpha_m=0.52$ is adopted), then with the results and Eq.(14), we
can obtain all the sizes of multiple giant components. Contrast
between theoretical and simulation is showed as follows:

\begin{table}[h]
\centering \caption{The fraction of components, $P_1$ and fraction
of the whole set $\mathcal S$ with $\alpha=1/2, 1/3, 1/4, 1/5$.
 The results are obtained by 100 random instances with $10^6$ nodes.
 The numbers in brackets are theoretical results, which are obtained
 by Eq.(14)-(17).} \label{table1}

\begin{tabular}{|c||c||c||c||c|}

\hline
 & $\alpha=1/2$ & $\alpha=1/3$ & $\alpha=1/4$ & $\alpha=1/5$ \\
\hline
$C_1$ & 0.5736 & 0.4142 & 0.3220 & 0.2631 \\
$C_2$ & 0.4144 & 0.3217 & 0.2629 & 0.2223 \\
$C_3$ &        & 0.2633 & 0.2223 & 0.1926 \\
$C_4$ &        &        & 0.1928 & 0.1699 \\
$C_5$ &        &        &        & 0.1519 \\
\hline $\sum C_i^2$ & 0.5007 & 0.3444 & 0.2594 &
0.2077\\
& (0.5000) & (0.3333) & (0.2500) & (0.2000) \\
\hline $x_m$ & 0.9880 & 0.9992 & 0.9999 &
0.9998\\
& (0.9802) & (0.9975) & (0.9997) & (0.9999) \\
\hline
\end{tabular}

\end{table}

In summary, parameter $\alpha$ determines when the system can
stabilize and which state the system can stabilize in. In the
evolution process of BFW($\alpha$) model, $\alpha$ can only take
effect on when to increase $k$. As to how much $k$ increases,
$\alpha$ doesn't work. With this special evolution rule of
BFW($\alpha$) model, the connection between two adjacent
steady-states is found and sizes of giant components are obtained.
As the value of $\alpha$ decreasing, theoretical results can be much
better verified by simulations.

\section{Conclusion and Discussion}

We detect the steady-state and evolution process of BFW($\alpha$)
model with both numerical and theoretical methods. According to the
rule of BFW($\alpha$) model, function $f_{\alpha}$ is defined to
calculate the change of the stage $k$ and number of giant components
$m$. Furthermore, by establishing the mathematical expression of
evolution equations on this model, an equal relationship between the
parameter $\alpha$ and steady-state condition is proved. Meanwhile,
with some hypothesis, the correspondence between parameter $\alpha$
and the number of giant components in steady-state is obtained, that
is when $\alpha\in(\frac{1}{m+1},\frac{1}{m}]$, BFW($\alpha$) model
must stabilize with $m$ giant components. Through the further
analysis of the evolution process and the numerical results, set
$\mathcal S$ is defined to find the rule of two components merging
before and after the threshold. Moreover, sizes of giant components
for different evolution strategies also has a close connection with
each others and satisfy some constraint equations, which is derived
from the evolution equations.

So far, we can calculate the number and sizes of giant components
for different evolution strategies with theoretical methods, which
can correspond with simulations very closely, especially when the
value of $\alpha$ is smaller than $0.25$. Additionally, the analysis
of the steady-state and evolution process is of great help to
explain why the percolation of BFW($\alpha$) model is explosive and
how explosive it is, which are almost supported by simulations
before. For example, before the the giant component appears, we can
obtain $\mathrm{d}k/\mathrm{d}t\sim O(k^{\frac{3}{2}}/n)$, although
the $k\sim O(n^{\frac{2}{3}+\delta})\sim o(n)$, for $0<\delta<1/3$,
we still have $\mathrm{d}k/\mathrm{d}t\sim\infty$; here $k\sim
C_{max}$, so the result means percolation of BFW($\alpha$) model is
explosive.

Besides, we just analyze this model on
Erd$\ddot{o}$s-R$\acute{e}$nyi random graph; as to other random
graph with any degree distribution, the theoretical methods in this
paper also work, which only need to modify the related probabilities
of the vertices of randomly sampled edge are (not) in the same
component.

\section*{Acknowledgment}

This work is supported by the Fundamental Research Funds for the Central Universities. \\

\end{document}